\documentstyle[11pt,psfig,twoside]{article}
\newlength{\defbaselineskip}
\newcommand{\setlinespacing}[1]%
           {\setlength{\baselineskip}{#1 \defbaselineskip}}
\newcommand{\onehalfspacing}{\setlength{\baselineskip}%
                           {1.5 \defbaselineskip}}

\newcommand{\be}{\begin{equation}}
\newcommand{\ee}{\end{equation}}
\newcommand{\bea}{\begin{eqnarray}}
\newcommand{\eea}{\end{eqnarray}}
\newcommand{\nno}{\nonumber}

\newcommand{\Bb}{{\bf B}}
\newcommand{\Eb}{{\bf E}}

\newcommand{\bx}{{\bf x}}
\newcommand{\Bbt}{{\bf \tilde{B}}}
\newcommand{\Ebt}{{\bf \tilde{E}}}

\newcommand{\ez}{{\bf \hat{e}_z}}
\newcommand{\pr}{{\partial}}
\newcommand{\n}{{\nabla}}
\newcommand{\al}{\alpha}
\newcommand{\bt}{\beta}
\newcommand{\sech}{\hbox{sech}}
\newcommand{\kp}{\kappa}

\begin{document}


\begin{flushright}
hep-th/0507210
\end{flushright}

\begin{center}
{\Large\bf Observable signals in a string inspired axion-dilaton background and 
Randall-Sundrum scenario}
\\[12mm]
Debaprasad Maity \footnote{E-mail: tpdm@iacs.res.in}, ~~Soumitra SenGupta \footnote{E-mail: 
tpssg@iacs.res.in}, ~ \\
{\em Department of Theoretical Physics, Indian Association for the
Cultivation of Science,\\ Calcutta - 700 032, India} \\
\bigskip

Sourav Sur \footnote{E-mail: sourav@iopb.res.in}\\
{\em Institute of Physics, Bhubaneswar - 751 005, India}\\[10mm]
\end{center}

\begin{abstract}
Rotation angle of the plane of polarization of the distant galactic radio waves 
has been estimated in a string inspired axion-dilaton background. It is found that the axion,dual to the field strength of the second 
rank massless antisymmetric Kalb-Ramond field in the string spectrum, produces 
a wavelength independent optical rotation which is much larger than that produced 
by the dilaton. Detection of such rotation has been reported in some recent 
cosmological experiments. The observed value has been compared with our estimated 
theoretical value following various cosmological constraints. The effects of warped 
extra dimensions in a braneworld scenario on such an optical rotation have been investigated.
\end{abstract}

\onehalfspacing

\section{{\large \bf Introduction} \label{intro}}
Despite several theoretical successes, string theory is yet to make any contact
with the observable world. The low energy field theory limit of a string theory
followed by suitable dimensional reduction yields the four dimensional action for 
the supergravity (SUGRA) multiplet coupled to super-Yang Mills. Some generic 
features of such an action may be explored for some indirect signal of a string 
inspired supergravity model. Two main areas to look for such signature of string 
theory are the cosmological/astrophysical experiments and high energy collider experiments.\\
This work aims to relate a characteristic feature of a string inspired model to
a much studied astrophysical phenomenon namely the rotation of the plane of 
polarization of distant galactic radiations\cite{masav}. It is well known,in the name of Faraday effect,
that the plane of polarization of a plane polarized electromagnetic wave rotates in
presence of a magnetic field. The rotation angle is directly proportional to the square of the
wavelength of the radiation. Extensive studies of this have been done in astrophysical\cite{brent} as well as 
other experiments\cite{zavat}. However some astrophysical data analysis\cite{ralston}revealed a small additional wavelength independent 
rotation of the plane of polarization over and above the usual Faraday rotation. 
A phenomenological model was immediately proposed\cite{cf} where 
a pseudo scalar field was coupled to the electromagnetic field for a possible explanation.
Such a model indeed can successfully explain such rotation of the plane of polarization of a plane polarized radiation. 
This possibility has also been  pointed out in several other works where various effects of 
axion( a pseudo-scalar)-electromagnetic coupling have been investigated in details\cite{taylor}.
In all these models the coupling term of the axion and electromagnetic field has a free coupling parameter
which can be fixed from experimental observations. 
It was however shown \cite{pmssg}in the context of a string inspired low energy models that such an 
axion-electromagnetic (em) coupling appears naturally from the consistency requirement like gauge anomaly
cancellation. Moreover the coupling parameter in such a theory gets determined from the theory itself.     
The axion-em coupling parameter in the higher dimensional theory is given by the inverse of Planck mass whereas the 
effective four dimensional coupling parameter in the four dimensional 
effective action depends on the type of compactification used to compactify the extra dimensions.   

The low energy four dimensional effective field theory action of string theory
\cite{gsw,pol,senrev}, namely the supergravity action, contains two massless fields, 
viz., a second rank antisymmetric tensor field (Kalb-Ramond field \cite{kr}) from 
the NS-NS sector of the underlying string theory as well as a scalar field called 
dilaton. The third rank field strength corresponding to the the Kalb-Ramond (KR) field
is dual to a pseudo scalar in four dimension called axion. Various consequences of the presence of such a axion  background in a 
curved spacetime on various physical phenomena have been investigated in 
view of possible indirect evidences of string theory at low energies\cite{ssg,others,hehl}. 
The interpretation of the KR field \cite{bmssg,pmssg} as a torsion in the background 
spacetime \cite{sabhehl}, inevitably implies a study of electromagnetism in torsion 
backgrounds. In this context, the gauge $U(1)$ Chern-Simons term that appears 
naturally on account of gauge anomaly-cancellation in the supergravity theory 
plays a crucial role in establishing a gauge-invariant coupling of the KR field 
(or, torsion) with the electromagnetic field \cite{pmssg,esposito}. Certain physically
observable phenomena, especially in the cosmological scenario, may result from 
the KR-electromagnetic coupling through the $U(1)$ Chern-Simons term. One such 
phenomenon of particular interest, argued to be induced by the KR field
\cite{skpmssgas,pdpj,preuss}, is a frequency-independent cosmic optical rotation 
of the plane of polarization of linearly polarized synchrotron radiation from 
high redshift galaxies \cite{birch}-\cite{ralston}. Other important aspects of 
gauge-invariant Einstein-KR-electromagnetic coupling include the effects on the 
electric-magnetic orthogonality, energy-conservation equation and also on the 
electric-magnetic duality symmetry \cite{skpmssgss}. \\
In this work we first take a four dimensional dilaton-axion-em coupled theory and estimate 
a theoretical bound on the value of the angle of rotation of the plane of polarization from known cosmological 
constraints. Subsequently we analyze this phenomenon in theories with extra dimensions.
Such theories  have drawn considerable attention both in the arena of high energy 
phenomenology as well as cosmology. Two of the most promising models are due 
to Arkani-Hamed, Dimopoulos and Dvali (ADD) \cite{add} and Randall and Sundrum (RS) 
\cite{rs}. Both these scenarios claim to solve the well known fine tuning problem 
of the Higgs mass against large radiative correction because of the gauge hierarchy. 
While the ADD model assumes the existence of large extra dimensions, the RS model 
proposes a four dimensional warped spacetime solution in a five dimensional bulk 
AdS spacetime.
Since both these models relate Planck scale 
to TeV scale , some signals of them are expected even in the forthcoming TeV scale 
collider. Our focus in this paper however centres on a cosmological scenario where even a tiny signal can grow to a 
measurable value over a large cosmological length scales.We therefore re-examine the phenomenon of the rotation of the plane of
polarization in presence extra compact dimensions. 
In particular we investigate this issue in 
the context of RS scenario where we show that the axion-electromagnetic 
coupling ( which depends on the compactification scheme used to compactify the extra dimensions)
changes drastically in the RS compactification resulting into some interesting consequences.

This paper addresses the above features as follows: in section \ref{compact},
we explore the effective four dimensional Einstein-Maxwell-Kalb-Ramond-dilaton
coupling from a five dimensional model. We study both the compactification
schemes, viz., Kaluza-Klein dimensional reduction on ${\mathbf S^1}$ and
RS type compactification on ${\mathbf S^1/Z_2}$. In section \ref{efffeq}, we
work out the effective field equations for the $U(1)$ gauge field as well
as the equations of motion for the KR axion (the hodge dual pseudo-scalar of the 
third rank antisymmetric KR field strength tensor in four dimensions) and the 
dilaton, in a cosmological scenario. The KR field induced phenomena of cosmic 
optical activity (leading to a birefringence in background spacetime) has been 
analyzed in detail in section \ref{optact}. We consider a homogeneous and 
isotropic cosmological background of Friedmann-Robertson-Walker (FRW) type 
and study the optical rotation in four dimensions when both dilaton and axion 
are present as 4+1 dimensional `bulk' fields. The $U(1)$ gauge field is also 
considered to be existing in the bulk. Precise estimates of the optical rotation 
angle is determined in the cases corresponding to both the compactification 
schemes discussed above. We also compute the maximum KR field energy density 
from the maximum bound on the optical rotation angle given by observational data 
for radiation from distant galaxies, and compare the result both with the Cosmic 
Microwave Background (CMB) energy density and with the present dark matter 
density of the Universe.

\section{{\large \bf Effective four dimensional Einstein-Maxwell-Kalb-Ramond-dilaton
theory from a five dimensional `bulk' model}  \label{compact}}

Let us consider a general five dimensional action for Einstein-Maxwell-Kalb-Ramond-dilaton 
(EMKRD) coupling in the Einstein frame
\be
S_{eff} ~=~ \int d^5 x \sqrt{- G} \left[ M^3 \left( R^{(5)} - \frac 1 2 \pr_M \Phi
\pr^M \Phi \right) -~ \frac 1 {12} e^{- 2 \Phi} H_{MNL} H^{MNL} ~-~ e^{- \Phi} F_{MN}
F^{MN} \right]    \label{action}
\ee
where $R^{(5)}$ is the five dimensional scalar curvature corresponding to the metric
$G_{MN}$; $M$ is the five dimensional
Planck mass; $\Phi$ is the massless dilaton field; $F_{MN} = \pr_{[M} A_{N]}$ is the
field strength of a $U(1)$ gauge field $A_N$; and $H_{MNL} = \pr_{[M} B_{NL]} + 2 
M^{- 3/2} A_{[M} F_{NL]}$ is the strength of the KR two form $B_{MN}$ augmented with 
$U(1)$ Chern-Simons term $A_{[M} F_{NL]}$. All the fields are supposed to live in the 
entire `bulk', i.e., in the usual 3+1 dimensional spacetime $(x)$ as well as the extra 
compact space $(y)$.

In order to have a four dimensional effective theory with Kalb-Ramond, dilaton and
the $U(1)$ gauge fields, we resort to two specific schemes of compactification as follows:

\subsection{{Scheme I : Kaluza-Klein dimensional reduction on ${\mathbf S^1}$} \label{sc1}}

In this type of compactification any field $\Psi (x,y)$ is decomposed in the form
\be
\Psi (x,y) ~=~ \frac 1 {\sqrt{L}} \sum_{n = 0}^{\infty} \psi^n (x) \exp\left[{\frac{2
\pi i n y} L}\right]
\ee
where $L$ is the radius of compactification. For the graviton, Kalb-Ramond and $U(1)$
gauge fields, various scalars and vector moduli fields appear by virtue of the above
scheme. The dilaton $\Phi$ however does not give rise to any such modular field. In fact,
it is convenient to shift the dilaton by a constant volume factor, so that in the 
effective four dimensional theory the dilaton is given by $\phi = \Phi - \ln V$. One 
can now write the four dimensional low energy effective action for the massless modes as
\be
S ~=~ \int d^4 x \sqrt{- g} \left[ M_p^2 \left( R - \frac 1 2 \pr_\mu \phi
\pr^\mu \phi \right) -~ \frac 1 {12} e^{- 2 \phi} H_{\mu\nu\lambda} H^{\mu\nu\lambda}
~-~ e^{- \phi} F_{\mu\nu} F^{\mu\nu} \right]    \label{action1}
\ee
where $R$ is the four dimensional scalar curvature corresponding to the four-metric
$g_{\mu\nu}$; the four dimensional Planck mass $M_p$ is related to the five dimensional
Planck mass $M$ as: ~ $M_p^2 = M^3 L$.
$H_{\mu\nu\lambda} = \pr_{[\mu} B_{\nu\lambda]} + 2 M_p^{-1} A_{[\mu} F_{\nu\lambda]}$ is
the KR field strength augmented with $U(1)$ Chern-Simons term. The massive modes, which 
have masses inversely related to the radius of the compact dimension, i.e. $L$, may be 
ignored as long as the radius is small $\sim M_p$, but can be significant in theories 
with large extra dimensions \cite{add}.

\subsection{{Scheme II : Compactification on ${\mathbf S^1/Z_2}$}  \label{sc2}}

Let us now resort to an alternative compactification scheme similar to that in
the Randall-Sundrum (RS) model. In the minimal version of such a model, described
in a five dimensional AdS spacetime, the extra coordinate $y$ is compactified on
a ${\mathbf S^1/Z_2}$ orbifold. Two branes, viz., the hidden and visible branes,
are located at two orbifold fixed points $y = 0$ and $y = \pi$ respectively. The
line element of the corresponding background
\be
ds^2 ~=~ e^{- 2 \sigma (y)} \eta_{\alpha\beta} dx^\alpha dx^\beta ~+~ r_c^2 dy^2
\ee
describes a non-factorisable geometry with an exponential warping over a flat
($\eta_{\alpha\beta}$) four dimensional part. The warp factor is given in
terms of the parameter $\sigma = \kp r_c |y|$, where $r_c$ is the compactification
radius and $\kp$ is of the order of the higher dimensional Planck scale $M$. The
four dimensional Planck mass $M_p$ is related to the five dimensional Planck mass
$M$ as:
$$M_p^2 = \frac{M^3} \kp \left(1 - e^{- 2 \kp r_c \pi}\right).$$

Starting from the five dimensional free action for the dilaton field $\Phi$:
\be
S_\Phi ~=~ - \frac {M^3} 2 \int d^5 x \sqrt{- G} ~G^{MN} \pr_M \Phi \pr_N \Phi
\ee
we apply  a Kaluza-Klein (KK) decomposition of the dimensionless dilaton field
$\Phi (x, y)$, viz.,
\be
\Phi (x, y) ~=~ \sum_{n = 0}^{\infty} \phi^{n} (x) ~ \zeta^n (y)
\ee
Now following the usual procedure of analyzing the bulk fields in RS scenario, 
we obtain the solution for the massive modes
\be
\zeta_n (y) ~\approx~ \left[ \frac{\sqrt{\kp r_c} e^{- \kp r_c \pi}}{J_3 \left(\frac{m_n} 
\kp e^{\kp r_c \pi}\right)}\right] e^{2 \kp r_c y} ~J_2 \left(\frac{m_n} \kp
e^{\kp r_c y}\right)  
\ee

For the zero mass mode ($n = 0, m_n = 0$), however, the solution of the equation
is given by
\be
\zeta_0 (y) ~=~  c_0 ~+~ \frac{c_1}{\kp r_c} e^{4 \kp r_c |y|} 
\ee
The condition of self-adjointness of course leads to $c_1 = 0$ and leaves the scope
of only a constant solution for $\zeta_0$. Using the normalization condition, one
finally obtains
\be
\zeta_0 ~=~ \sqrt{\kp r_c} ~e^{- \kp r_c \pi}
\ee
which indicates that the massless mode of the dilaton field is suppressed by
a huge warp factor on the visible brane. The situation is similar to the RS
compactification of the free Kalb-Ramond Lagrangian, for which the massless
KR projection is suppressed by the same exponential factor \cite{bmsenssgPRL}.
The massive modes, however, have different solutions in this case.

As we are interested in a low energy effective four dimensional compactified
version of the Einstein-Maxwell-KR-dilaton theory, we take into account only
the massless modes of the projections of all the bulk fields on the visible
brane. Now, in the bulk the dilaton has exponential couplings with both the KR
field and the $U (1)$ gauge field. Therefore, when one considers only the
massless dilaton mode which carries an enormous suppression factor as shown above,
it is clear that the exponential dilaton couplings can be very well approximated as 
$e^{-\phi} \sim 1$, in the four dimensional effective action on the visible brane. 
In other words, restricting one's attention to the effects at low energies, we
are in a sense led to a scenario in which the Einstein-Maxwell-KR Lagrangian
is supplemented only with the free dilaton Lagrangian $\frac 1 2 (\partial\phi)^2$
and there is virtually no exponential dilaton coupling with either the KR field or 
the $U(1)$ gauge field. The RS compactification of the free Kalb-Ramond Lagrangian 
has already been studied extensively in \cite{bmsenssgPRL}. Similar kind of 
compactification for the bulk $U (1)$ gauge field has been studied in \cite{dhr} 
and also elucidated in \cite{dmssg} in the context of cosmic optical activity. We 
write down the essential results obtained therein as follows:

Decomposing the bulk KR field $B_{\mu\nu} (x, y)$ and the bulk $U (1)$ gauge field
$A_\mu (x, y)$ respectively as
\be
B_{\mu\nu} (x, y) ~=~ \frac 1 {\sqrt{r_c}} \sum_{n = 0}^{\infty} B_{\mu\nu}^n (x)
\chi^n (y) ~;~~~~~~~~~~~~
A_{\mu} (x, y) ~=~ \frac 1 {\sqrt{r_c}} \sum_{n = 0}^{\infty} A_{\mu}^n (x)
\psi^n (y)
\ee
and performing an analysis similar to that shown above for the dilaton, one finds
the solution for the massless mode of the KR field \cite{bmsenssgPRL} and that
for the $U (1)$ gauge field \cite{dhr} as
\be
\chi^0 ~=~ \sqrt{\kp r_c} e^{- \kp r_c \pi} ~;~~~~~~~~~~~~~~
\psi^0 ~=~ \frac 1 {\sqrt{2 \pi}} .
\ee

Finally, taking everything together and remembering that the three-tensor $H_{MNL}$
in the bulk is given by the strength of the KR field $B_{MN}$ augmented with the
$U (1)$ Chern-Simons term $A_{[M} F_{NL]}$, one obtains the RS compactified four
dimensional low energy effective EMKRD action as
\be
S ~=~ \int d^4 x \sqrt{- g} \left[ M_p^2 \left( R ~-~ \frac 1 2 \pr_\mu \phi
\pr^\mu \phi \right) -~ \frac 1 {12} e^{- 2 \zeta_0 \phi} H_{\mu\nu\lambda} H^{\mu\nu\lambda}
~-~ e^{- \zeta_0 \phi} F_{\mu\nu} F^{\mu\nu} \right]    \label{action2}
\ee
where $\zeta_0 = \sqrt{\kp r_c} e^{- \kp r_c \pi} \sim 10^{- 17}$ for $\kp r_c = 12$ 
(chosen to solve the naturalness problem in the standard model), and we denote
the massless mode $\phi^0$ of the dilaton as $\phi$ which we call the massless four 
dimensional dilaton field. Similarly we also omit the superscript $``0"$ in the massless 
modes $B_{\mu\nu}^0 (x)$ and $A_{\mu}^0 (x)$ for the KR field and the $U (1)$ gauge 
field respectively. The three tensor $H_{\mu\nu\lambda}$ is now given by
\be
H_{\mu\nu\lambda} ~=~ \pr_{[\mu} B_{\nu\lambda]} ~+~ 2 \sqrt{\frac{\kp}{M_p^3}}
e^{\kp r_c \pi} A_{[\mu} F_{\nu \lambda]}
\ee
which shows a huge enhancement ($\sim e^{\kp r_c \pi}$) in the coupling between the KR 
field strength (dual to the string axion) and the $U(1)$ gauge field because of the 
Chern-Simons term. This enhancement is a major outcome of the RS compactification 
implemented on the KR-Maxwell part of the five dimensional action (\ref{action}) as has 
been shown explicitly in \cite{dmssg}.

\section{{\large \bf Effective Field equations in a cosmological scenario}  \label{efffeq}}

In presence of standard cosmological matter, for example perfect fluid, we write down
a general four dimensional low energy effective EMKRD action as
\be
S ~=~ \int d^4 x \sqrt{- g} \left[ M_p^2 \left( R ~-~ \frac 1 2 \pr_\mu \phi
\pr^\mu \phi \right) -~ \frac 1 {12} e^{- 2 \al \phi} H_{\mu\nu\lambda} H^{\mu\nu\lambda}
~-~ e^{- \al \phi} F_{\mu\nu} F^{\mu\nu} ~+~ {\mathcal L}_m \right]    \label{actiong}
\ee
where
\be
H_{\mu\nu\lambda} ~=~ \pr_{[\mu} B_{\nu\lambda]} ~+~ \frac {2 \bt} {M_p} A_{[\mu}
F_{\nu \lambda]} ~;            \label{krg}
\ee
is the modified KR field strength, ~${\mathcal L}_m$~ is the cosmological matter Lagrangian
density, and $\al$ and $\bt$ are two free dimensionless parameters  whose values are specified
according to the two schemes of compactification respectively as:
\bea
&& \al ~=~ \bt ~=~ 1 ~: \hskip 2.8in  \hbox{Scheme I} \nno \\
&& \al ~=~ \sqrt{\kp r_c} e^{- \kp r_c \pi} \ll 1 ,~ \bt ~=~ \sqrt{\frac{\kp}{M_p}} e^{\kp r_c 
\pi} \gg 1 ~:~~~~~~~~~~~~~ \hbox{Scheme II} .  \label{values}
\eea

In terms of the massless and dimensionless pseudo scalar `axion' field $\xi$, which 
is related to $~H^{\mu\nu\lambda}~$ via the duality
\be
H^{\mu\nu\lambda} ~=~ M_p ~e^{2 \al \phi}~ \epsilon^{\mu\nu\lambda\rho} ~\pr_{\rho}
\xi ~,         \label{axion}
\ee
the modified Maxwell (electromagnetic) equations are given by
\bea
&(a)&~~~ \n_{\mu} \left(e^{- \al \phi} F^{\mu\nu} ~+~ \bt~ \xi 
~^* F^{\mu\nu}\right) =~ 0 \nno\\
&(b)&~~~ \n_{\mu}~^* F^{\mu\nu} ~=~ 0        \label{maxeq}
\eea
where  $\n_{\alpha}$ stands for the covariant derivative and $~^*F_{\alpha\beta}
= \epsilon_{\alpha\beta\mu\nu} F^{\mu\nu}$ is the Hodge dual of the electromagnetic
field strength $F^{\mu\nu}$.

The coupled equations of motion for the axion and dilaton are expressed as
\bea
&(a)&~~~ \n_{\mu} \left(e^{2 \al \phi} \pr^{\mu} \xi\right) =~ \frac \bt {M_p^2}
~ F_{\mu\nu} ~^* F^{\mu\nu} \nno\\
&(b)&~~~ \n_{\mu} \pr^{\mu} \phi ~=~ \al ~e^{2 \al \phi}~ \pr_{\mu} \xi \pr^{\mu} \xi ~-~
\frac \al {M_p^2} ~e^{- \al \phi} ~F_{\mu\nu} F^{\mu\nu} .       \label{dilax}
\eea

We consider the spatially flat homogeneous and isotropic
Robertson-Walker line element,
\be
d s^2 ~=~ a^2 (\eta) \left[- d \eta^2 ~+~ d {\bf x}^2\right]       \label{rw}
\ee
where $a (\eta)$ is the scale factor; and the conformal time $\eta$ is related to the cosmic
time $t$ through the relation ~$dt = a (\eta) d\eta$.

The electromagnetic field equations take the form
\bea
&(a)&~~~~~ \n  \cdot \Ebt ~=~ \al~ \n \phi \cdot \Ebt ~-~
\bt ~ e^{\al \phi}~ \n \xi \cdot \Bbt \nno \\
&(b)&~~~~~ \dot{\Ebt} ~-~ \nabla \times \Bbt ~=~ \al \left(\dot{\phi}~\Ebt
~-~ \n \phi \times \Bbt\right) ~-~ \bt~ e^{\al \phi} \left(\dot{\xi} \Bbt ~+~ \nabla 
\xi \times \Ebt \right) \nno\\
&(c)&~~~~~ \nabla \cdot \Bbt ~=~ 0 \nno \\
&(d)&~~~~~ \dot{\Bbt} ~+~ \nabla \times \Ebt ~=~ 0  \label{genemfeq}
\eea
where the overhead dot stands for partial differentiation with respect to $\eta$,
and we denote
\be
\Ebt (\eta, \bx) ~\equiv~ a^2 (\eta) ~\Eb (\eta, \bx)~;~~~~~~
\Bbt (\eta, \bx) ~\equiv~ a^2 (\eta) ~\Bb (\eta, \bx)
\ee
$\Eb$ and $\Bb$ being the usual electric and magnetic fields respectively.

The equations of motion for the axion and the dilaton reduce to
\bea
&(a)&~~~ \ddot{\xi} ~+~ 2 \frac{\dot{a}} a \dot{\xi} ~-~ \n^2 \xi ~+~ 2 \al \left(\dot{\phi}
~\dot{\xi} ~-~ \n \phi \cdot \n \xi\right) ~=~ \frac{2 \bt }
{M_p^2 ~a^4} e^{- 2 \al \phi}  \Ebt \cdot \Bbt \nno \\
&(b)&~~~  \ddot{\phi} ~+~ 2 \frac{\dot{a}} a \dot{\phi} ~-~ \n^2 \phi
~-~ \al ~ e^{2 \al \phi} \left(\dot{\xi}^2 ~-~  |\n \xi|^2 \right) ~=~ \frac{2 \al}
{M_p^2 ~a^4} e^{- \al \phi} \left(\Ebt^2 ~-~ \Bbt^2\right) .               \label{gendilax}
\eea

It follows from the electromagnetic field equations (\ref{genemfeq}) that
\be
\pr_\eta \left(e^{- a \phi} \Ebt \cdot \Bbt\right) =~ e^{- a \phi} \left(\Bbt \cdot \n 
\times \Bbt ~-~ \Ebt \cdot \n \times \Ebt\right) -~ \bt \left(\dot{\xi}~ |\Bbt|^2 ~+~ 
\n \xi \cdot \Ebt \times \Bbt\right)
\ee
which shows that the vectors $\Ebt$ and $\Bbt$ are \underline{not} in general mutually
orthogonal.  This is in sharp contrast with the plane wave nature of the free electric and
magnetic fields ~$(\Ebt, ~\Bbt ~\sim ~e^{i (k \eta - {\bf k \cdot x})})$~ for $\phi = \xi 
= 0$. This leads to an optical anisotropy (and/or an effective magnetization) in the 
background spacetime in presence of the dilaton and the axion, which is manifested 
observationally through a cosmic optical rotation of the plane of polarization of 
radiation from distant galactic sources. Such an optical activity, which makes a 
hitherto isotropic medium birefringent, is in fact entirely due to the typical 
axionic coupling with the electromagnetic field through the $U (1)$ Chern-Simons 
term as shown in \cite{skpmssgas,skpmssgss}. The dilaton on its own does not produce the
birefringence, however, there may be a substantial change in the amount of axion-induced
optical rotation in presence of the dilaton. In the following section we study this optical
activity in presence both axion and dilaton and determine the amount of optical rotation
of the plane of polarization of galactic radiation in both flat background and spatially flat
Friedmann-Robertson-Walker (FRW) background.

Over the large cosmological scales we take into account, in our subsequent analysis,
only the temporal variations of the axion and dilaton fields, i.e., $\xi = \xi (\eta) , 
~ \phi = \phi (\eta)$, whence the equations (\ref{gendilax}) take the forms (on dropping
the ${\cal O} (M_p^{- 2})$ terms)
\bea
&(a)&~~~ \pr_\eta \left(a^2 ~e^{2 \al \phi}~ \dot{\xi}\right) ~=~ 0  \nno\\
&(b)&~~~ \pr_\eta \left(a^2 ~\dot{\phi}\right) ~=~ \al ~ e^{2 \al \phi} ~a^2 ~\dot{\xi}^2
\eea
These can be solved exactly to obtain
\be
\xi (\eta) ~=~ \frac 1 {\al} \tanh \left[\al ~h \int^\eta  \frac{d \eta'}{a^2 (\eta')}\right] ~;~~~~~~~
\phi (\eta) ~=~ \frac 1 {\al}  \ln \left\{\cosh \left[\al ~h \int^\eta
\frac{d \eta'}{a^2 (\eta')}\right]\right\}          \label{dilaxsol}
\ee
where the constant parameter $h$ determines the rates of evolution of the
free axion and dilaton fields.

\section{{\large \bf Cosmic Optical Activity in presence of axion and
dilaton}                 \label{optact}}

The electromagnetic wave equations in presence of the KR axion and the dilaton are 
given by
\bea
&(a)&~~~ \Box \Bbt ~\equiv~ \ddot{\Bbt} ~-~ \n^2 \Bbt
~=~ \al~ \dot{\phi}~\dot{\Bbt} ~+~ \bt ~e^{\al \phi}~ \dot{\xi}~ \n \times \Bbt   \nno\\
&(b)&~~~ \Box \Ebt ~\equiv~ \ddot{\Ebt} ~-~ \n^2 \Ebt ~=~ \al \left(\dot{\phi}~\dot{\Ebt}
~+~ \ddot{\phi}~\Ebt\right) +~ \bt ~e^{\al \phi} \left[\dot{\xi} ~\n \times \Ebt ~- 
\left(\ddot{\xi} ~+~ \al \dot{\phi} \dot{\xi}\right) \Bbt\right] .    \label{emweq}
\eea

We assume general wave solutions of the form
\be
\Bbt (\eta, {\bf x}) ~=~  \Bbt_0 (\eta) ~ e^{- i {\bf k \cdot x}} ~;~~~~~~~~~~
\Ebt (\eta, {\bf x}) ~=~  \Ebt_0 (\eta) ~ e^{- i {\bf k \cdot x}} .    \label{emwsol}
\ee
Taking the $z$-direction as the propagation direction of the electromagnetic waves, i.e.,
${\bf k} = k \ez$, the equations for the polarization states, viz.,
\be
b_{\pm} (\eta) ~=~ \tilde{B}_{0x} (\eta) ~\pm~ i~ \tilde{B}_{0y} (\eta) ~;~~~~~~~~~
e_{\pm} (\eta) ~=~ \tilde{E}_{0x} (\eta) ~\pm~ i~ \tilde{E}_{0y} (\eta) 
\ee
can be rewritten as
\bea
&(a)&~~~ \ddot{b}_{\pm} ~-~ \al ~\dot{\phi} ~\dot{b}_{\pm} ~+ \left(k^2 ~\mp~
k \bt e^{\al \phi} \dot{\xi}\right) b_{\pm} ~=~ 0 \nno \\
&(b)&~~~ \ddot{e}_{\pm} ~-~ \al ~\dot{\phi} ~\dot{e}_{\pm} ~-~ \al ~\ddot{\phi}
~ e_\pm ~+ \left(k^2 ~\mp~ k \bt e^{\al \phi} \dot{\xi}\right) e_{\pm} ~=~
- \bt ~ e^{\al \phi} \left( \al \dot{\phi} \dot{\xi} ~+~ \ddot{\xi}
\right) b_\pm .    \label{pol}
\eea

Now, to obtain the amount of rotation of the plane of polarization  of electromagnetic
radiation, given by the difference between the phases of the $b_+$ and $b_-$ waves (or
between those of the $e_+$ and $e_-$ waves), we solve the above equations (\ref{pol}) 
separately for the different values of the parameters $\al$ and $\bt$ corresponding to 
the compactification schemes I and II described in section \ref{compact}. In what follows, 
we make an estimate of the optical rotation angle for both the schemes, first in a 
flat spacetime background and followed by a spatially flat background of FRW type.

\subsection{Flat spacetime background                     \label{flat}}

In order to make a preliminary idea as to how the coupling of KR axion and also the dilaton
with Einstein-Maxwell theory affects the electromagnetic waves, and thereby results into 
optical activity in the radiation coming from distant galactic sources, we first consider the
simplest situation --- that is, of a flat Universe with cosmological scale factor ~
$a = 1$~ and ~$\eta$~ identified as the usual (cosmic) time coordinate $t$. 

\vskip .2in
\noindent
{\bf Case I : ~$\al = 1 ,~ \bt = 1$}
\vskip .1in

In this case, which corresponds to the four-dimensional effective theory obtained through the
Kaluza-Klein dimensional reduction of a higher dimensional EMKRD action, the solutions
(\ref{dilaxsol}) for the axion and dilaton fields take the form:
\be
\xi (t) ~=~ \tanh \left(h t\right) ~;~~~~~~~~
\phi (t) ~=~ \ln \left[\cosh \left(h t\right)\right]          \label{dilaxflat}
\ee
while the equations (\ref{pol}) for the polarization states reduce to
\bea
&(a)&~ \ddot{b}_{\pm} ~-~ h ~\tanh (h t) ~\dot{b}_{\pm} ~+
\left[k^2 ~\mp~ k h ~\sech (h t)\right] b_{\pm} ~=~ 0 \nno \\
&(b)&~ \ddot{e}_{\pm} ~-~ h ~\tanh (h t) ~\dot{e}_{\pm} ~-~
h^2 \sech^2 (h t) ~ e_\pm ~+ \left[k^2 ~\mp~ k h ~\sech (h t)\right] e_{\pm}
~=~ h^2 \sech (h t) \tanh (h t)~ b_\pm .   \nno\\   \label{polflat1}
\eea

In general, it is very difficult to solve these equations exactly. However, while
studying the KR field induced cosmic electromagnetic effects and the influence of the
dilaton, we consider that the KR axion and the dilaton perturb the Maxwell's equations
by a small amount in a standard FRW cosmological background with both matter and
radiation dominant. Thus the KR field as well as the dilaton is assumed to have a
negligible effect on shaping the cosmological background spacetime. One way of
thinking about this is to imagine that the KR axion and the dilaton decouple from
the radiation or matter (dust) fluid, shaping the cosmic geometry, far prior to 
dust-photon decoupling, leaving behind a `cosmic axion-dilaton background' which 
affects incoming radiation from distant galaxies albeit rather softly. The electromagnetic 
effects produced as such will no doubt gradually subside, as the Universe expands further, 
however, may yet  be observable in this epoch. In view of the smallness of such 
electromagnetic effects induced in axion-dilaton background, we consider the parameter 
$h$, which has the dimension of inverse time, to be sufficiently smaller than the wave 
number $k$ of the electromagnetic waves.  Assuming the standard phase-exponential ansatz 
of WKB type for the polarization states $b_{\pm}$ and $e_{\pm}$:
\bea
&(a)&~~ b_{\pm} (t) ~=~ b_0~ e^{i k S_b^{\pm} (t)} ~;~~~~~~~~~
S_b^{\pm} (t) = S_{b0}^{\pm} (t) + \frac{S_{b1}^{\pm} (t)} k +
\frac{S_{b2}^{\pm} (t)} {k^2} + \cdots ~,       \nno\\
&(b)&~~ e_{\pm} (t) ~=~ e_0~ e^{i k S_e^{\pm} (t)} ~;~~~~~~~~~
S_e^{\pm} (t) = S_{e0}^{\pm} (t) + \frac{S_{e1}^{\pm} (t)} k +
\frac{S_{e2}^{\pm} (t)} {k^2} + \cdots
\eea
with constant amplitudes $b_0$ and $e_0$, we obtain
\be
S_b^{\pm} (t) ~\approx~ S_e^{\pm} (t) ~=~ t ~-~ \frac 1 k \left[ \frac i 2 ~
\ln \left\{\cosh (h t)\right\} ~\pm~ \tan^{- 1} \left\{\tanh\left(\frac{h t} 2\right)\right\}\right]
+ {\cal O} \left(\frac 1 {k^2}\right)      \label{polsol}
\ee
Dropping the ${\cal O} (k^{- 2})$ terms in the above expression, we finally obtain
\be
b_\pm (t) ~=~ b_0 \sqrt{\cosh (h t)} ~ \exp \left[i k \left(t ~\mp~ \frac 1 k
\tan^{- 1} \left\{\tanh\left(\frac{h t} 2\right)\right\}\right)\right]
\ee
and a similar expression for $e_\pm$. Thus, in presence of the KR axion and the dilaton
there is a modulation of the amplitude as well as an alteration of the phase of free
electromagnetic waves even in a flat spacetime background. The measure of the
optical rotation angle can be obtained from the phase difference
\be
\Delta (t_F) ~=~ \vert\arg f_+ ~-~ \arg f_-\vert_{t = t_F}
~=~ 2 \tan^{- 1} \left[\tanh \left(\frac{h t_F} 2\right)\right]           \label{rotflat1a}
\ee
where ~$f_\pm$~ is either ~$b_\pm$~ or ~$e_\pm$ ,~ and  $t_F$ is the time interval
between the emission and reception of the electromagnetic radiation in a flat background.
For $h t_F < 1$ the above expression can be expanded as
\be
\Delta (t_F) ~=~ h t_F ~+~ \frac{(h t_F)^3} 6 ~+~ {\cal O} (h t_F)^5 .       \label{rotflat1}
\ee

Let us now compare the results obtained here with that obtained in ref.\cite{skpmssgas}
in which the effect of the axion (and not the dilaton) on the electromagnetic waves has 
been studied. In a flat spacetime background, the axion has been found to produce
an optical rotation given by the amount $h t_F$ upto ${\cal O} (k^{- 2})$, where $h$
determines the rate of evolution of the KR axion \cite{skpmssgas}. As is evident from
the expression (\ref{rotflat1a}) or (\ref{rotflat1}), the dilaton produces an additional
rotation given by an amount at most ~$\sim h^3 t_F^3$ ~(for $h t_F < 1$). Moreover,
the amplitude modulation of the free electromagnetic waves discussed above is entirely
due to the dilaton as the axion alone cannot produce such an effect.

\vskip .2in
\noindent
{\bf Case II : ~$\al  = \sqrt{\kp r_c} e^{- \kp r_c \pi} \ll 1 ,~ \bt = \sqrt{\frac{\kp}{m_p}}
e^{\kp r_c \pi} \gg 1$}
\vskip .1in

In this case, which corresponds to the four-dimensional effective theory obtained
via RS compactification of a higher dimensional EMKRD action, the exponential
dilaton coupling with the KR axion as well as with the electromagnetic field can 
be approximated to be $\sim 1$ as the parameter $\alpha \sim 10^{-16}$ for $\kp r_c = 12$. 
On the other hand, the KR--electromagnetic coupling is enormously enhanced through the 
parameter $\bt$ which is $\sim 10^{15}$ for $\kp r_c = 12$. We consider the value of 
the parameter $\al$ to be zero for all practical purposes, whence in a flat background 
the solution for the axion is simply given by ~$\xi = h t$ ,~ and the equations for 
the polarization states take the form
\be
\ddot{f}_{\pm} ~+ \left(k^2 ~\mp~ k \bt h\right) f_{\pm} ~=~ 0     \label{polflat2}
\ee
where ~$f_\pm$~ is either ~$b_\pm$~ or ~$e_\pm$. The solution of this equation
is given exactly as
\be
f_\pm ~\sim~ \exp \left(i~ \sqrt{k^2 ~\mp~ k \bt h}~ t\right) .
\ee
which implies that the angle of rotation of the plane of polarization of the electromagnetic
waves given by
\be
\Delta (t_F) ~=~ \vert\arg f_+ ~-~ \arg f_-\vert_{t = t_F} ~=~ k t_F \left(\sqrt{1
~+~ \frac {\bt h} k} ~-~ \sqrt{1 ~-~ \frac {\bt h} k}\right)  .                     \label{rotflat2a}
\ee

Now, for $\kp r_c = 12$, the value of $k/\bt = 2 \pi e^{- \kp r_c \pi} c/\lambda$ ~(where
$\lambda$ is the wavelength of the incoming radiation) which is typically $\sim 10^{- 25}$ 
GeV for visible radiation and $\sim 10^{- 31}$ GeV for radio waves. For $h < k/\bt$, the 
above expression for the optical rotation angle is given by
\bea
\Delta (t_F) ~=~  \bt h t_F ~+~ (\bt h)^2 \frac{t_F}{8 k} ~+~ \cdots  .    \label{rotflat2}
\eea

\subsection{Spatially flat FRW spacetime background                               \label{spflat}}

Let us now consider a spatially flat expanding FRW background spacetime. Setting
the present value of the cosmological scale factor to unity, one can express the latter 
in terms of the redshift $z$ as
\be
a ~=~ \frac 1 {(1 ~+~ z)} .
\ee
The flat space expressions for the optical rotation angle $\Delta$ given
in the previous subsection \ref{flat}, for the two cases I and II, can be generalized
in a spatially flat Universe by replacing the time interval $t_F$ in Eqs.(\ref{rotflat1a})
and (\ref{rotflat2a}) by the cosmic lookback time $t_L$ which is given in terms
of redshift $z$ as
\be
t_L (z) ~=~ \int_0^z \frac{d z'}{(1 + z') ~H (z')}                \label{lbdef}
\ee
where $H = a^{- 1} (d a/d t)$ is the Hubble parameter. In a spatially flat FRW 
Universe, the lookback time $t_L$ is the same as the distance traveled along light 
path from source to observer. 

In what follows, we consider the standard spatially flat cosmological model where 
both (baryonic plus dark) matter and dark energy are dominant. The Hubble parameter is 
expressed as
\be
H ~=~ H_0 \sqrt{\frac{\Omega_{m} ~+~ \Omega_{d}}{a^3} ~+~ \Omega_{X}}
\ee
where $H_0$ is the Hubble constant ($\sim 1.7 \times 10^{- 18}$ s$^{- 1}$); and 
$\Omega_{b}, \Omega_{d}$ and $\Omega_{X}$ are the present values of the density 
parameters for the ordinary baryonic matter, dark matter and dark energy respectively. 
The total density parameter $\Omega = \rho/\rho_c = \Omega_{b} + \Omega_{d} + 
\Omega_{\Lambda} = 1$, where $\rho$ is the total matter-energy density of the Universe 
and $\rho_c = 3 H_0^2/(8 \pi G)$ is critical density ($\sim 5 \times 10^{- 27}$ Kg 
m$^{- 3}$). The present accepted values of $\Omega_{b}, \Omega_{d}$ and $\Omega_{X}$
are close to $0.05, 0.25$ and $0.7$ respectively. The expression (\ref{lbdef}) for 
the lookback time reduces to
\be
t_L (z) ~=~ \frac 1 {3 H_0 \Omega_X} \left[ \ln \left(\frac{\sqrt{(\Omega_b + 
\Omega_d) (1 + z)^3 ~+~ \Omega_X} ~-~ \Omega_X}{\sqrt{(\Omega_b + 
\Omega_d) (1 + z)^3 ~+~ \Omega_X} ~+~ \Omega_X}\right) -~ \ln \left(\frac{1 ~-~
\Omega_X}{1 ~+~ \Omega_X}\right) \right]  .              \label{lb}
\ee
We estimate the value of the optical rotation angle $\Delta$ resorting
separately to the two cases I and II corresponding to the two schemes of
compactification and ascertain the maximum limit on the parameter $h$ and hence
on the effective KR field energy density $\rho_{KR}$ from the observational bounds
on the optical rotation angle \cite{cf,cfj,ralston}.

Catalogues of observational data for radio galaxies and quasars \cite{burspin}
at distances comparable to the Hubble length ~($ = 1/H_0 \sim 10^{10}$ y) have
been analyzed in great detail in \cite{cf,cfj} for redshift $z \geq 0.4$. The
observable of interest is the angular separation ~$|\chi - \psi|$ ,~ where $\psi$
is the orientation angle of the axis of the radio source; and $\chi$ is the
observed polarization angle extracted from the data on separating out the usual
effect of Faraday rotation that occurs due to the passage of radiation through
galactic/inter-galactic magnetized plasma. The Faraday rotation effect, which is
proportional to the inverse square of the wave number $k$ of the incoming radiation,
is removed from the data by making a straight line fit of the total polarization
angle $\theta$ as: ~ $\theta = f k^{- 2} + \chi$, where $f$ is a Faraday rotation
measure that depends on the plasma density and the parallel component of the
background magnetic field along the direction of propagation of the wave. We
refer to the upper bound on the observed amount of $k$-independent
optical rotation angle obtained in \cite{cfj} by analyzing the data for values of $|\chi
- \psi|$ in the range $45^o - 135^o$: ~
\be
\Delta^{(obs)}_{max} ~=~ 6^o~~~~~~~ \hbox{at redshift}~~z = 0.4 ~.        \label{obsrot}
\ee

\vskip .2in
\noindent
{\bf Case I : ~$\al = 1 ,~ \bt = 1$}
\vskip .1in

On replacing the time interval $t_F$ by the lookback time $t_L (z)$ ~[Eq.(\ref{lb}] in 
the expression (\ref{rotflat1a}) for $\Delta$ (obtained in the flat space analysis for 
the case $\al = \bt = 1$),we have the optical rotation angle for the spatially flat 
Universe given by
\be
\Delta (z) ~=~ 2 \tan^{- 1} \left[\tanh \left(\frac {h t_L (z)} 2\right)\right] \label{rotmd1}
\ee
with $t_L (z)$ as given in Eq.(\ref{lb}). Using the above upper bound ($\sim 6^o$) on 
$\Delta$ at $z = 0.4$, the maximum upper limit on the parameter $h$ can be obtained: ~ 
$h_{max}^{(obs)} = 5 \times 10^{- 42}$ GeV .~ This implies that the effective KR field 
energy density which is given by
\be
\rho_{KR} ~=~ \frac{M_p^2}{2 a^2} ~e^{4 \phi} ~\dot{\xi}^2 ~=~  \frac{M_p^2 h^2 (1 + z)^2} 2
\ee
is observationally limited to a maximum value  ~$\rho_{KR}^{max} = 2 \times 10^{- 46}$ 
GeV$^4$ ~ at $z = 0.4$. Here we have used the relation $a = (1 + z)^{-1}$ and the solutions 
(\ref{dilaxflat}), corresponding to the case $\al = \bt = 1$,  for the axion and dilaton 
fields. This value of $\rho_{KR}^{max}$ is sufficiently larger than the Cosmic Microwave 
Background (CMB) energy density ~$\rho_{cmb} \sim 4 \times 10^{- 52}$ GeV$^4$ ~ (corresponding 
to the temperature $T_{cmb} = 2.7$ K) but only two order larger than the dark matter energy 
density ~$\rho_d = \rho_c \Omega_d \sim 1.6 \times 10^{- 48}$ GeV$^4$ . Lack of direct 
experimental support in favour of such a high background KR energy density, however, imposes 
a severe constraint on the value of $\rho_{KR}^{max}$. If we take  $\rho_{KR}^{max}$ to 
be at most of the order of $\rho_{cmb}$, then the maximum value of the parameter $h$ is 
practically reduced to  ~$h_{max}^{(cmb)} = 6.7 \times 10^{- 45}$ GeV, which implies a 
substantial reduction in the maximum optical rotation angle $\Delta \sim 25$ arcsec at 
$z = 0.4$. On the other hand, if $\rho_{KR}^{max}$ is taken to be $\sim \rho_{dm}$, then 
the maximum value of $h$ is $h_{max}^{(dm)} = 1.3 \times 10^{- 43}$ GeV, which implies a 
fairly larger maximum limit on $\Delta \sim 0.2^o$. In the second case one may hope for 
a possible measurement of such an effect with the improvement of observational accuracy. 

It should now be mentioned here that for all the three upper bounds on $h$, viz., 
~$h_{max}^{(obs)} \sim 10^{- 42}$ GeV, $h_{max}^{(dm)} \sim 10^{- 43}$ GeV and 
$h_{max}^{(cmb)} \sim 10^{- 45}$ GeV, the contribution of the dilaton on the amount of 
optical rotation is extremely small, being at most cubic in $h t_L (z)$. For $h = 
h_{max}^{(obs)}$, the maximum change produced on $\Delta$ by the dilaton is $\delta\Delta 
\sim 0.05^o$, for $h = h_{max}^{(dm)}$, the dilaton's contribution is extremely 
insignificant $\delta\Delta \sim 0.003$ arcsec, while for $h = h_{max}^{(cmb)}$, 
the dilaton practically inflicts no change.

\vskip .2in
\noindent
{\bf Case II : ~$\al  = \sqrt{\kp r_c} e^{- \kp r_c \pi} \ll 1 ,~ \bt = \sqrt{\frac{\kp}{m_p}}
e^{\kp r_c \pi} \gg 1$}
\vskip .1in

Substituting the time interval $t_F$ by the lookback time $t_L (z)$ in the flat space 
expression (\ref{rotflat2a}) for $\Delta$, the optical rotation angle for the spatially 
flat FRW Universe can be given in terms of the redshift $z$ as
\be
\Delta (z) ~=~ k \left(\sqrt{1 ~+~ \frac {\bt h} k}
~-~ \sqrt{1 ~-~ \frac {\bt h} k}\right) t_L (z)                            \label{rotmd2}
\ee
where $t_L (z)$ is as given in Eq.(\ref{lb}). 

The maximum observational bound for $\Delta$ [Eq.(\ref{obsrot})] sets a new upper limit
on the parameter $h$ at $z = 0.4$: ~ $h_{max}^{(obs)} = 2 \times 10^{- 58}$ GeV .~ As
such at $z = 0.4$, the effective KR field energy density is now limited to a extremely
low maximum value ~$\rho_{KR}^{max} = 4 \times 10^{- 79}$ GeV$^4$.This implies an 
unnatural fine-tuning of the KR energy density to a incredibly small value in order 
to avoid a catastrophically large wavelength independent rotation of the plane of 
polarization hitherto unobserved. 

\section{{\large \bf Conclusion}  \label{conclu}}

In this paper we have shown that a string inspired SUGRA model 
can offer a possible explanation for the hitherto unexplained wavelength independent
optical rotation observed in the distant galactic radio waves.
Among various possible sources one of the possible source of such a rotation is known to be the coupling between 
a pseudo-scalar and electromagnetic field. Such a coupling term was originally proposed from a 
phenomenological 
point of view by Carroll {\it et al} \cite{cf,cfj} and considered also by 
\cite{ralston,sikivie,raffelt,jain}.However the coupling parameter in these models 
were arbitrary and needed to be determined either from experimental data or from some more
fundamental theoretical framework where such a term occurs naturally.  
It turned out that the gauge invariant axion-em coupling originating 
in the context of anomaly cancellation in a string based SUGRA 
model exactly yields the desired coupling term. However the exact value of the coupling parameter depends   
crucially on the type of compactification used to compactify the extra dimensions.We have first shown 
that the other scalar field in such models, namely the dilaton, doesn't play any significant 
role in the phenomenon of optical rotation. Analysing the usual dilaton-axion-em coupled theory in four dimension ,
we found that despite a low value of the KR energy density, we can get 
a near measurable value of the optical rotation angle because of the large cosmological 
distance through which the em wave travels across the space. We have estimated an upper 
bound of this KR axion induced rotation by comparing the KR energy density with the known CMB and dark matter energy density.
This whole analysis is subsequently carried out in the context of a braneworld scenario.
The value of the coupling parameter in effective four dimensional theory , which depends on the type of compactification,
gets enhanced by a large warp factor in RS scenario.This results into 
an unacceptably large value of the rotation of the plane of polarization perhaps indicating  
an inherent conflict between string inspired models and Randall-Sundrum scenario. It has been 
shown in a recent work \cite{acpm} that a similar enhancement takes place for 
gravitational waves also.  Some more exhaustive studies to explore this apparent conflict, 
are currently being carried out by the authors. It can be easily seen by extending our analysis that in the 
case of ADD compactification, no such large enhancement of the KR-electromagnetic coupling 
would take place which may lead to a high value of the rotation angle as found in RS 
scenario. Thus from this cosmological observation, it appears that ADD compactification is 
more compatible with string inspired models than the RS compactification.  
This work therefore proposes a possibility to find a signature of the string inspired low 
energy model in astrophysical experiments. 

\vskip .2in
\noindent
{\bf {\Large Acknowledgment}}
\vskip .1in

DM acknowledges the Council of Scientific and Industrial Research, Govt. of India for
providing financial support. SS acknowledges Department of Atomic Energy, Govt. of 
India for financial support.

\end{document}